\documentclass[aps,prl,twocolumn,superscriptaddress]{revtex4}
\usepackage{subfigure}
\usepackage{graphicx}
\usepackage{rotating} \usepackage{color}
 \usepackage{amsmath,amsthm}
\usepackage{epstopdf}
\begin{document}
\title{Universal relations in the self-assembly of proteins and RNA}
\author{D. Thirumalai}
\affiliation{Institute for Physical Science and Technology, University of Maryland, College Park, Maryland 20742, USA}

\begin{abstract}
Concepts rooted in physics are becoming increasingly important in biology as we transition to an era in which quantitative descriptions of all processes from molecular to cellular level are needed. In this essay I discuss two unexpected findings of universal behavior, uncommon in biology, in the self-assembly of proteins and RNA. These findings, which are surprising, reveal  that physics ideas applied to biological problems ranging from folding to gene expression to cellular movement and communication between cells might lead to discovery of  universal principles operating in adoptable living systems.
\end{abstract}
\maketitle


The fascination of physicists and mathematicians with biology can be traced back to at least the turn of the twentieth century. Two remarkable books, one on morphogenesis grappling with patterns on animals and plants by Thompson \cite{Thompson} and the other by Schrodinger \cite{Schrodinger} outlining how the principles of statistical mechanics and stochastic fluctuations govern life itself, illustrate beautifully the need for a quantitative perspective on biology. These works, which have inspired many scientists to delve into the complexities of biology, continue to be relevant to this day. The need for molecular explanations of a number of phenomena in biology also began early in the 1900s when Christian Bohr, the father of Niels Bohr, discovered the impact of pH (Bohr effect) on oxygen binding to hemoglobin, a problem that has dominated research in molecular biophysics. Although the use of physics ideas in biology is not new the intensity with which physical principles are being applied to living systems is unprecedented, drawing researchers from a variety of backgrounds. As a result, there is hardly any subfield of biology in which the presence of physics ideas is not felt. 

My interest in using the tools of physics to understand biology began in the late 1980s when I got interested in protein folding \cite{Honeycutt90PNAS}. Dana Honeycutt and I had worked on the effects of randomly placed obstacles on the shapes of homopolymers, a prelude to my interests in crowding effects on proteins and RNA that came years later.  After we finished that work, Dana asked me if I had any suggestions on what we could do next, and he proposed that we tackle the dynamics of a polymer molecule in a random environment. I talked him out of this because the problem was (and is) difficult.  I also felt that the results might be of interest to only a small number of scientists. Instead, I suggested that we look into protein folding. It is generally assumed, thanks to the groundbreaking experiments by Anfinsen \cite{Anfinsen73Science}, that the number of folded states of a protein is indeed small (in fact unique) implying that it is likely that the functionally competent state corresponds to the lowest (or near lowest) free energy state amidst all the exponentially large number of conformations a polypeptide chain can adopt.  I wanted to understand why proteins have to reach the lowest free energy minimum (Anfinsen hypothesis) in order to execute their functions.  I theorized to Dana that proteins  could have countable number of low free energy minima in which they could adopt similar structures without compromising their ability to carry out their functions. In other words, a functioning protein could be metastable.  Using a minimal model of a protein, one that has been adapted and modified by many researchers, Honeycutt and I \cite{Honeycutt90PNAS} argued that the functional state of a protein could be metastable, a notion which was not inconsistent with experiments but nevertheless was considered heretical. Upon publication of our paper \cite{Honeycutt90PNAS}, Kaufmann wrote us that he too had similar ideas and discovered that biologists were not too receptive to them. He did alert us to a series of experimental papers reporting that functional states (plural!) could be metastable. Interestingly, more recently it has been demonstrated that the folded state of mammalian prions may well be metastable \cite{Baskakov01JBC} with the more stable conformation being aggregation prone, and hence deleterious. The concept that folded states could be metastable seems more easily accepted for ribozymes \cite{Solomatin10Nature}. 

Prior to the announcement of the metastability hypothesis in \cite{Honeycutt90PNAS}, papers demonstrating   uses of  statistical mechanics of disordered systems \cite{Bryngelson89JPC} and polymer theory \cite{Dill85Biochem} as a way to describe the self-assembly of proteins  appeared. The modern perspectives have literally transformed the field of protein folding research, a trend that continues unabated to this day. Thus, time was ripe to produce a fresh perspective on the self-assembly of proteins and RNA.  We were the first to use physics  concepts to describe the complex pathways in the folding of RNA \cite{Thirumalai96ACR}.  Our initial work got me hooked and my research group has since then focussed on using physics concepts to describe a number of problems in biology. Here, I will give two example where we unexpectedly discovered universal relations in protein and RNA folding using ideas rooted in polymer physics and glasses \cite{Kirkpatrick14RMP}, which at least to me is a surprise given the tremendous emphasis on specificity placed by biologists. 

{\bf Length dependence on folding cooperativity and collapse transition:} Single domain proteins undergo a remarkably cooperative transition from the unfolded to a folded state when the conditions for folding become favorable, for example by lowering the temperature. The folding reaction, occurring at  $T_F$ may be viewed as a phase transition. Single domain proteins are finite-sized, with the number of amino acid residues in the majority of the experimentally characterized proteins containing less than about 100 residues. Thus, we expect finite-size effects to play an important role in the folding phase transition. Because proteins are polymers, with water being a poor solvent for these systems, we expect that in the process of folding proteins should also undergo a collapse transition, at a characteristic temperature, $T_{\theta}$. Based on theoretical arguments and precise numerical results for protein-like models we established sometime ago that efficient folding occurs if $ T_{\theta} \approx T_F$ \cite{Camacho93PRL}, a prediction that has only recently been fully validated \cite{Hofmann12PNAS}. The extent of cooperativity in this transition can be assessed using a dimensionless quantity, $\Omega_c = \frac{T_F^2}{\Delta T}|\frac{df_{NBA}}{dT}|$ where $f_{NBA}$ is the fraction of molecules in the NBA at $T$, $\Delta T$ is the full width at half maximum of $|\frac{df_{NBA}}{dT}|$. In the expression for $\Omega_c$, $|\frac{df_{NBA}}{dT}|$ is evaluated at $T_F$. We showed that $\Omega_c \sim N^{\zeta}$ \cite{Li04PRL} where $N$ is the number of amino acid residues in a protein and $\zeta = 1 + \gamma$ with $\gamma$ being the exponent that characterizes the divergence
of the susceptibility at the critical point for a $n$-component
ferromagnet with $n=0$, corresponding to  a self-avoiding walk - a reasonable model that describes the global properties of  unfolded proteins. An accurate numerical estimate based on a fifth order $\epsilon$ expansion of $\phi^4$ field theory for a polymer gives $\gamma \approx 1.22$, leading to the conclusion that $\Omega_c \sim N^{1.22}$. We showed \cite{Li04PRL} that experiments on a number of proteins are in accord with this {\it universal} prediction. 

The rationale for expecting the universal behavior for $\Omega_c$ goes as follows: (1) By analogy with magnetic systems we can identify $f_{NBA}$ as an order parameter that distinguishes between the folded and unfolded states, and hence $T\frac{df_{NBA}}{dT}$ can be associated with  "susceptibility" with $T$ being the ordering field. (2) Camacho and I \cite{Camacho93PNAS} showed that the collapse transition at $T_{\theta}$ for finite $N$ could be second order while the folding transition at $T_F$ is (weakly) first order. The condition $T_{\theta} \approx T_F$ implies that the folding transition itself could have tricritical character, thus tidily explaining the marginal stability of proteins. Therefore, the critical exponents that control the behavior of the polypeptide chain at $T_{\theta}$ should manifest themselves in the folding phase transition. Given that susceptibility scales as $N^{\gamma}$ and $\frac{\Delta T}{T_F} \sim N^{-1}$ \cite{Li04PRL}, it follows that $\Omega_c \approx N^{\zeta}$. 

{\bf Folding rates of proteins and RNA scale as  $e^{{\sqrt N}}$:} Using theoretical arguments, whose genesis is in the dynamics of activated transitions in supercooled liquids \cite{Kirkpatrick14RMP}, I suggested that folding rates of proteins can be written as,
$k_F = k_0 \exp{(-\alpha N^{\beta})}$ where $\beta$ should be 0.5 and $\alpha$ is a constant on the order of unity \cite{Thirum95JPI}. Similar arguments also lead to the same scaling for RNA folding as well \cite{Hyeon12BJ}. The essence of the argument for expecting a sub-linear dependence of the barrier height on $N$ can be understood by noting that the driving force for folding proteins is to bury the hydrophobic residues whereas charged or polar residues are better accommodated by extending the chain. This intrinsic conflict produces some fraction of interactions that are favorable for folding and others that favor extended structures.  Similarly, in RNA  
favorable base-pairing interactions and the  hydrophobic nature of the bases tend to collapse RNA whereas the charged phosphate residues prefer extended structures. Thus, the distribution of activation free energy, $\Delta G^{\ddagger}_{UF}/k_BT$,  between the folded and unfolded states is a sum of favorable and unfavorable terms. We expect from central limit theorem that the distribution of $\Delta G^{\ddagger}_{UF}/k_BT$ should be roughly Gaussian with dispersion $\langle (\Delta G^{\ddagger}_{UF})^2\rangle\sim N$.   Thus,  $\Delta G^{\ddagger}_{UF}/k_BT\sim N^{\beta}$  with $\beta=1/2$.  

  \begin{figure}[]
\includegraphics[width=3.00in]{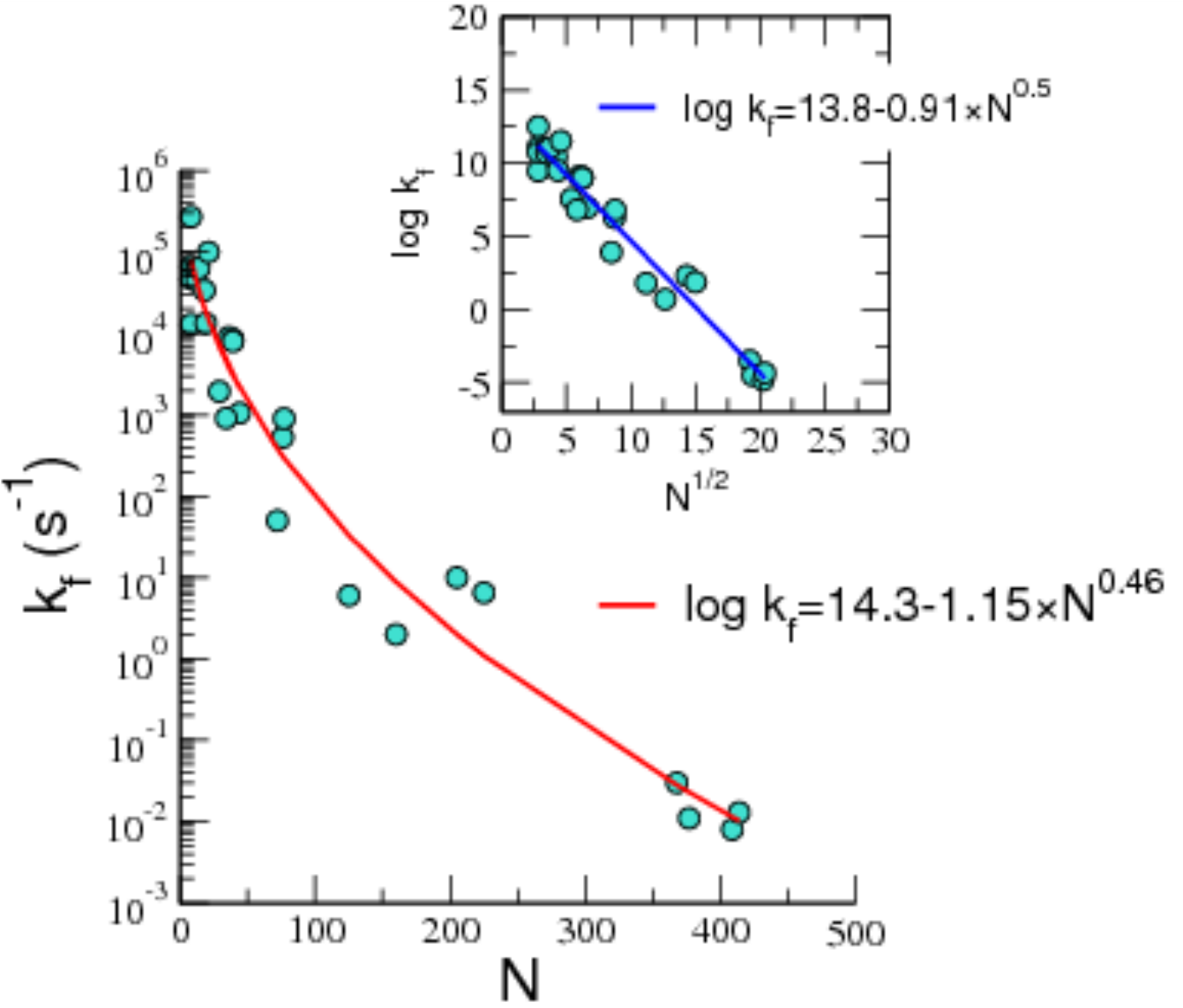}
  \caption{Dependence of the folding rate of RNA on $N$. The solid line is the theoretical fit with $\beta \approx 0.46$ as a free parameter. The inset on top shows that the quality of the fit ,over seven orders of magnitude, is excellent with $\beta$ fixed at the theoretical value of 0.50. The figure is adapted from \cite{Hyeon12BJ}.}
\end{figure}

Remarkably, for both proteins (see Fig. 7 in ref. \cite{Dill11PNAS}) and RNA (Fig. 1) the predicted dependence of the folding rate on $N$ is extremely well-described by the theoretical prediction. From the fits of theory to experiments, we find that that the inverse of the prefactor for RNA , $k_0^{-1}=\tau_0\approx 0.87$ $\mu s$, is almost six orders of magnitude larger than the transition state theory estimate of $h/k_BT\approx 0.16$ $ps$. 
The value of $\tau_0$, which  coincides with the typical base pairing time \cite{porschke1973BP}, is hence the speed limit for RNA folding. The predicted value for $\tau_0$ is close to the speed limit established for protein folding as well \cite{Eaton04COSB}. The common speed limit suggests that the initial events triggering folding (in all likelihood favorable loop formation leading to hairpin formation in RNA and nucleation in protein folding) may be similar. 

{\bf Final Remarks:} The illustrations here do not even come close to capturing the excitement of working at the interface between biology and physics. The prospects of unearthing general principles governing living systems using physics concepts have never been greater. As a result the ever evolving biological world is a perfect playground for physicists. As the applications grow more quantitative, it is natural that physicists will come to grips with crucial differences between the living and non-living matter. The notions of adaptation and evolution, which play a crucial role in living matter at all length scales, have to be integrated into the theoretical description of biological processes. It is clear that that in grappling with these problems, physicists, who are most definitely up to the task, will perpetually feel like kids in a candy store! 

{\bf Acknowledgements:} I am grateful to an extraordinary group of talented scientists who taught me a great deal during their involvement in biophysics research. These researchers, many of whom are in academic institutions world wide, are most responsible for my continued excitement in biophysics research. I am grateful to Krastan Blagoev for inviting me to write this piece and for useful comments Our works have been generously supported by the National Science Foundation for nearly twenty five years most recently through CHE 13-61946.


\end{document}